\PassOptionsToPackage{unicode}{hyperref}
\PassOptionsToPackage{hyphens}{url}
\PassOptionsToPackage{dvipsnames,svgnames,x11names}{xcolor}
\documentclass[
]{article}
\usepackage{amsmath,amssymb}
\usepackage{lmodern}
\usepackage{iftex}
\ifPDFTeX
  \usepackage[T1]{fontenc}
  \usepackage[utf8]{inputenc}
  \usepackage{textcomp} 
\else 
  \usepackage{unicode-math}
  \defaultfontfeatures{Scale=MatchLowercase}
  \defaultfontfeatures[\rmfamily]{Ligatures=TeX,Scale=1}
\fi
\IfFileExists{upquote.sty}{\usepackage{upquote}}{}
\IfFileExists{microtype.sty}{
  \usepackage[]{microtype}
  \UseMicrotypeSet[protrusion]{basicmath} 
}{}
\makeatletter
\@ifundefined{KOMAClassName}{
  \IfFileExists{parskip.sty}{%
    \usepackage{parskip}
  }{
    \setlength{\parindent}{0pt}
    \setlength{\parskip}{6pt plus 2pt minus 1pt}}
}{
  \KOMAoptions{parskip=half}}
\makeatother
\usepackage{xcolor}
\usepackage{color}
\usepackage{fancyvrb}

\DefineVerbatimEnvironment{Highlighting}{Verbatim}{commandchars=\\\{\}}
\newenvironment{Shaded}{}{}

\newcommand{\DecValTok}[1]{\textcolor[rgb]{0.25,0.63,0.44}{#1}}

\newcommand{\NormalTok}[1]{#1}
\newcommand{\OperatorTok}[1]{\textcolor[rgb]{0.40,0.40,0.40}{#1}}

\usepackage{graphicx}
\makeatletter
\def\maxwidth{\ifdim\Gin@nat@width>\linewidth\linewidth\else\Gin@nat@width\fi}
\def\maxheight{\ifdim\Gin@nat@height>\textheight\textheight\else\Gin@nat@height\fi}
\makeatother
\setkeys{Gin}{width=\maxwidth,height=\maxheight,keepaspectratio}
\makeatletter
\def\fps@figure{htbp}
\makeatother
\providecommand{\tightlist}{%
  \setlength{\itemsep}{0pt}\setlength{\parskip}{0pt}}
\newlength{\cslhangindent}
\newlength{\csllabelwidth}
\newlength{\cslentryspacingunit} 
\newenvironment{CSLReferences}[2] 
 {
  \setlength{\parindent}{0pt}
  \ifodd #1
  \let\oldpar\par
  \def\par{\hangindent=\cslhangindent\oldpar}
  \fi
  \setlength{\parskip}{#2\cslentryspacingunit}
 }%
 {}
\usepackage{calc}

\ifLuaTeX
\usepackage[bidi=basic]{babel}
\else
\usepackage[bidi=default]{babel}
\fi
\babelprovide[main,import]{american}

\def\languageshorthands#1{}
\ifLuaTeX
  \usepackage{selnolig}  
\fi
\IfFileExists{bookmark.sty}{\usepackage{bookmark}}{\usepackage{hyperref}}
\IfFileExists{xurl.sty}{\usepackage{xurl}}{} 
\hypersetup{
  pdftitle={Spelunker: A quick-look Python pipeline for JWST NIRISS FGS
Guide Star Data},
  pdfauthor={Derod Deal, Néstor Espinoza},
  pdflang={en-US},
  colorlinks=true,
  linkcolor={Maroon},
  filecolor={Maroon},
  citecolor={Blue},
  urlcolor={Blue},
  pdfcreator={LaTeX via pandoc}}

\title{Spelunker: A quick-look Python pipeline for JWST NIRISS FGS Guide
Star Data}

\usepackage[affil-it]{authblk}
\usepackage{orcidlink}
\author[1%
  ]{Derod Deal%
    \,\orcidlink{0009-0006-6758-4751}\,%
    }
\author[2%
  ]{Néstor Espinoza%
    \,\orcidlink{0000-0001-9513-1449}\,%
    }

\affil[1]{Department of Astronomy, University of Florida P.O. Box
112055, Gainesville, FL, USA}
\affil[2]{Space Telescope Science Institute, 3700 San Martin Drive,
Baltimore, MD 21218, USA}
\date{14 November 2023}

\begin{document}
\maketitle

\hypertarget{summary}{%
\section{Summary}\label{summary}}

\texttt{spelunker} is a Python library that provides several tools for
analyzing and visualizing JWST NIRISS FGS guide star data products. The
library can download guide star data from the Mikulski Archive for Space
Telescopes (\protect\hyperlink{ref-marston_overview_2018}{Marston et
al., 2018}) for any given Program ID in a single line of code. Through
an efficient parallelization process, the pipeline can use this data to
extract photometry in seconds and point-spread function information from
all frames within a selected Program ID in minutes. Data loaded within
the package allows for many possible analyses such as Gaussian fitting,
periodograms of each fitted Gaussian parameter, flux time series, and
photometry optimization.

In addition, \texttt{spelunker} provides visualization and analysis
tools to study these products in detail, including the incorporation of
JWST engineering telemetry, which can be used to put the JWST FGS data
products in context with other observatory variables that might help
explain data patterns both in the primary science data products and in
the JWST FGS data itself. The library also cross-references, tracks and
stores guide star metadata for the user. This metadata includes
information such as the GAIA ID of the guide star, coordinates, and
magnitudes, among others. This pipeline empowers users to study guide
star data with applications like technical anomaly detection and
time-domain astronomy at a high 64 ms cadence. Existing pipeline such as
\texttt{jwst} provides the structure for calibrataing NIRISS/FGS,
NIRSpec, NIRCam, and MIRI data, such as with JWST \texttt{Datamodels}.
\texttt{spelunker} tools are engineered to explore FGS guide stars in
two main categories: as a method to detect technical anomalies in JWST
data, and as a pipeline to study time-domain astronomy with guide stars.

\hypertarget{statement-of-need}{%
\section{Statement of need}\label{statement-of-need}}

The James Webb Space Telescope
(\protect\hyperlink{ref-gardner_james_2023}{Gardner et al., 2023})
produces some of the highest sensitivity imaging of the cosmos across
all instruments. One of them, the NIRISS Fine Guidance Sensor
(\protect\hyperlink{ref-doyon_jwst_2012}{Doyon et al., 2012}), provides
guide star imaging with a passband of 0.6 to 5 microns through two
separate channels, each with a \(2.3’ \times 2.3’\) field of view (FOV)
and a sampling rate of 64 ms---data that is taken in parallel and is
thus available for every JWST observing program. While the onboard
system uses guide stars to provide information to the attitude control
system (ACS) which stabilizes the observatory, the astronomical
community can also use the data products associated with these 64 ms
cadence images as science products. Usages range from studying guide
star photometry in search of transient phenomena to using these data to
identify and investigate technical anomalies that might occur during
scientific observations with the rest of the JWST instruments. Despite
this wide range of possible usages, these data products are not
straightforward to manipulate and analyze, and there is no publicly
available package to download, investigate, and research guide star
data. \texttt{spelunker} is a Python library that was developed to
enable access to these guide star data products and their analysis.

\begin{figure}
\centering
\includegraphics{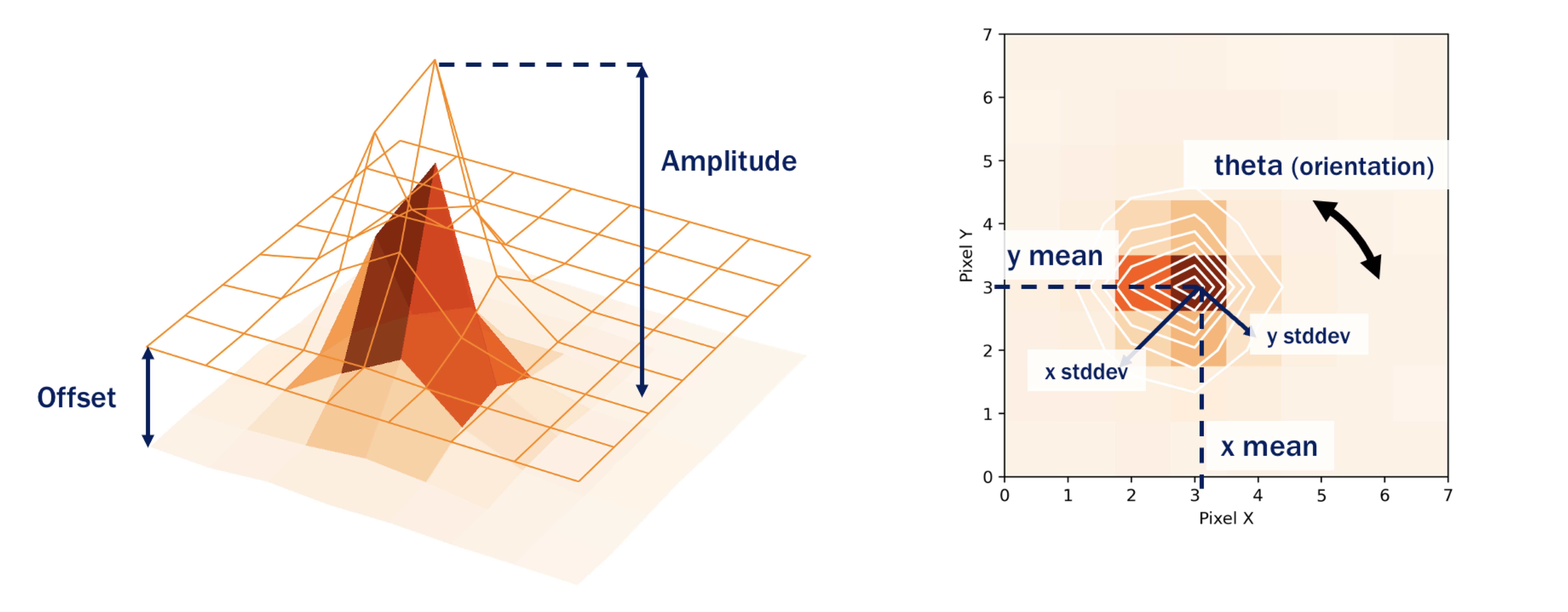}
\caption{There are seven parameters \texttt{gauss2d\_fit} measures:
amplitude (counts of the guide star), x pixel coordinate, y pixel
coordinate, the x and y standard deviations, theta (orientation of the
Gaussian model), and the offset (the background counts). This diagram
visualizes what each parameter represents on the Gaussian model.
\label{fig:Gaussian_diagram}}
\end{figure}

\hypertarget{overview-of-spelunker}{%
\section{Overview of Spelunker}\label{overview-of-spelunker}}

\texttt{spelunker} allows anyone to download and utilize fine guidence
guide star data from the JWST. Users can use the following lines of code
to download GS-FG data into an object:

\begin{Shaded}
\begin{Highlighting}[]
\NormalTok{spk }\OperatorTok{=}\NormalTok{ spelunker.load(pid}\OperatorTok{=}\DecValTok{1534}\NormalTok{)}
\end{Highlighting}
\end{Shaded}

\begin{figure}
\centering
\includegraphics{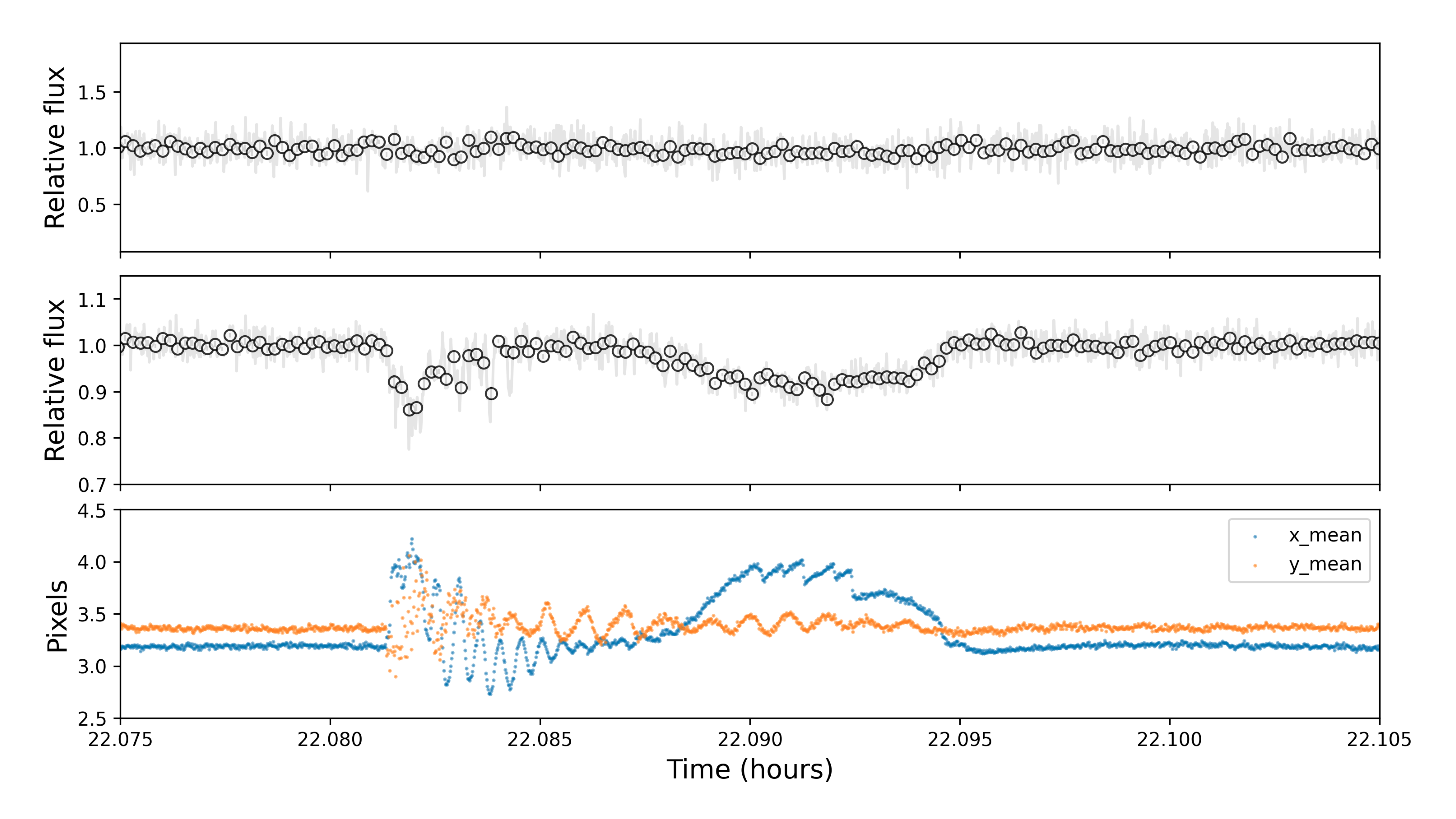}
\caption{A snippet from the guide star time series from Cycle 1 GO
Program ID 1803, observation 1 and visit 1. \textbf{Top} --- The guide
star time series of PID 1803 after loading it into \texttt{spelunker}
using \texttt{timeseries\_binned\_plot}. The time series uses the sum of
counts in each guide star fine guidance (GS-FG) frame. The data has no
significant features. \textbf{Middle} --- The same time series after
applying \texttt{optimize\_photometry} to the guide star light curve.
There are now prominent drops in the flux, which were previously unseen
with the raw time series data. \textbf{Bottom} --- Gaussian fitted x
pixel coordinate and y pixel coordinate for each frame in this section
of time series data. The guide star shifts around in this time series,
highlighting the core function of the ACS. \label{fig:guidestar_1803}}
\end{figure}

\texttt{spelunker} uses \texttt{astroquery} and MAST
(\protect\hyperlink{ref-marston_overview_2018}{Marston et al., 2018}) to
find and download GS-FG FITS files. There are several functions that can
manipulate and analyze guide star data:

\begin{itemize}
\tightlist
\item
  \textbf{\texttt{gauss2d\_fit}} A spatial Gaussian fit is applied to
  each of the frames loaded into the \texttt{spk} object. The Gaussian
  will fit the amplitude, pixel coordinates, pixel standard deviations,
  the model orientation, and the background offset. A diagram of
  Gaussian fitting parameters is shown in
  \autoref{fig:Gaussian_diagram}. The Gaussian measurements are then
  stored in \texttt{spk.gaussfit\_results} as an astropy table. Fitting
  spatial Gaussians to your guide star data can allow you to reveal
  technical anomalies that might be confused with science data from
  NIRISS or other JWST instruments and within the guide star flux time
  series (see \autoref{fig:guidestar_1803}).
\end{itemize}

\begin{itemize}
\item
  \textbf{\texttt{mnemonics}} Users can access JWST engineering
  telemetry and mnemonics using the \texttt{mnemonics} function. With a
  MAST API token, any mnemonic is accessible. High-gain antenna (HGA)
  movement and NIRISS filter wheel current are two examples of events
  that overplot science data to identify technical events on the
  telescope. Anomalies detection in guide star data or data from NIRISS,
  NIRCAM, NIRSpec, and MIRI is one of the primary capabilities of this
  function. Using \texttt{mnemonics} requires the additional
  installation of the \texttt{jwstuser} library.
\item
  \textbf{\texttt{periodogram}} This function uses the Lomb-Scarle
  periodogram (\protect\hyperlink{ref-lomb_least-squares_1976}{Lomb,
  1976}; \protect\hyperlink{ref-scargle_studies_1982}{Scargle, 1982}) to
  detect periodicities in guide star Gaussian fits. Periods in Gaussian
  fitted parameters like x and y pixel coordinates highlight systematics
  for an entire PID.
\item
  \textbf{\texttt{optimize\_photometry}} \texttt{optimize\_photometry}
  extracts the highest SNR pixels to optimize raw guide star photometry
  loaded from \texttt{spelunker}. \autoref{fig:guidestar_1803}
  demonstrates that \texttt{optimize\_photometry} reveals more
  information from guide star time series than that produced by the sum
  of counts in each frame.
\end{itemize}

With the mentioned tools, \texttt{spelunker} utilizes object oriented
programming (OOP) to store handy variables and its outputs, for
instance, 1D and 2D time series, guide star time arrays, and JWST data
models. Running \texttt{gauss2d\_fit}, \texttt{periodogram}, and
\texttt{mnemonics} will store their outputs in accessible attributes.
Useful properties of the guide star are stored in these attributes (for
instance, guide star galactic coordinates, GAIA ID, and stellar
magnitudes).

There are various plotting and visualization tools integrated into
\texttt{spelunker}'s workflow. One useful function is
\texttt{timeseries\_binned\_plot}, which automatically plots a binned
time series. The functions \texttt{gauss2d\_fit}, \texttt{periodogram},
and \texttt{mnemonics} have \texttt{matplotlib} axes returned for
straightforward plotting. Animations of spatial time series are another
visualization tool covered under \texttt{spelunker}. Of course, users
have the option to use the guide star data and results from attributes
to generate plots.

\hypertarget{acknowledgements}{%
\section{Acknowledgements}\label{acknowledgements}}

We would like to thank the Space Telescope Science Institute and the
National Astronomy Consortium for the opportunity to develop this
project. In particular, we acknowledge funding and support from the 2023
version of the Space Astronomy Summer Program (SASP) at STScI that made
it possible for the authors to work together on this project.

\hypertarget{references}{%
\section*{References}\label{references}}
\addcontentsline{toc}{section}{References}

\hypertarget{refs}{}
\begin{CSLReferences}{1}{0}
\leavevmode\vadjust pre{\hypertarget{ref-doyon_jwst_2012}{}}%
Doyon, R., Hutchings, J. B., Beaulieu, M., Albert, L., Lafrenière, D.,
Willott, C., Touahri, D., Rowlands, N., Maszkiewicz, M., Fullerton, A.
W., Volk, K., Martel, A. R., Chayer, P., Sivaramakrishnan, A., Abraham,
R., Ferrarese, L., Jayawardhana, R., Johnstone, D., Meyer, M., \ldots{}
Sawicki, M. (2012). {The JWST Fine Guidance Sensor (FGS) and
Near-Infrared Imager and Slitless Spectrograph (NIRISS)}. In M. C.
Clampin, G. G. Fazio, H. A. MacEwen, \& Jr. Oschmann Jacobus M. (Eds.),
\emph{Space telescopes and instrumentation 2012: Optical, infrared, and
millimeter wave} (Vol. 8442, p. 84422R).
\url{https://doi.org/10.1117/12.926578}

\leavevmode\vadjust pre{\hypertarget{ref-gardner_james_2023}{}}%
Gardner, J. P., Mather, J. C., Abbott, R., Abell, J. S., Abernathy, M.,
Abney, F. E., Abraham, J. G., Abraham, R., Abul-Huda, Y. M., Acton, S.,
Adams, C. K., Adams, E., Adler, D. S., Adriaensen, M., Aguilar, J. A.,
Ahmed, M., Ahmed, N. S., Ahmed, T., Albat, R., \ldots{} Zondag, E.
(2023). The {James} {Webb} {Space} {Telescope} {Mission}.
\emph{Publications of the Astronomical Society of the Pacific},
\emph{135}, 068001. \url{https://doi.org/10.1088/1538-3873/acd1b5}

\leavevmode\vadjust pre{\hypertarget{ref-lomb_least-squares_1976}{}}%
Lomb, N. R. (1976). Least-{Squares} {Frequency} {Analysis} of
{Unequally} {Spaced} {Data}. \emph{Astrophysics and Space Science},
\emph{39}, 447--462. \url{https://doi.org/10.1007/BF00648343}

\leavevmode\vadjust pre{\hypertarget{ref-marston_overview_2018}{}}%
Marston, A., Hargis, J., Levay, K., Forshay, P., Mullally, S., \& Shaw,
R. (2018). {Overview of the Mikulski Archive for space telescopes for
the James Webb Space Telescope data archiving}. \emph{Observatory
Operations: Strategies, Processes, and Systems VII}, \emph{10704},
1070413. \url{https://doi.org/10.1117/12.2311973}

\leavevmode\vadjust pre{\hypertarget{ref-scargle_studies_1982}{}}%
Scargle, J. D. (1982). Studies in astronomical time series analysis.
{II}. {Statistical} aspects of spectral analysis of unevenly spaced
data. \emph{The Astrophysical Journal}, \emph{263}, 835--853.
\url{https://doi.org/10.1086/160554}

\end{CSLReferences}

\end{document}